\documentclass[]{aa}

\usepackage[dvipsnames]{xcolor}
\usepackage{float}

\def\rmit#1{{\it #1}}              
\def\specchar#1{{\sc #1}}

\def\SiI{\mbox{Si\,\specchar{i}}}

\def\CaII{\mbox{Ca\,\specchar{ii}}}

\def\eg{\rmit{e.g.}}

\usepackage{booktabs}
\usepackage{graphicx}
\usepackage{multirow}
\usepackage[normalem]{ulem}
\usepackage{color}
\usepackage[flushleft]{threeparttable}

\usepackage{array}
\newcolumntype{?}{@{\vrule width 2pt}}

\usepackage{hyperref}
\hypersetup{
    final=true,
    pageanchor=true,
    colorlinks=true,
    breaklinks=true,
    linkcolor=blue,
    citecolor=blue,
    urlcolor=blue,
    pdfpagemode=UseNone,
    pdftitle={},
    pdfauthor={T. Felipe et al.},
    pdfsubject={Solar Physics},
    pdfkeywords={Methods: observational -- Sun: chromosphere -- Sun: oscillations  -- sunspots -- Techniques: polarimetric}}

\titlerunning{Opacity effects in NLTE inversions}   

\begin{document}



\title{Impact of opacity effects on chromospheric oscillations inferred from NLTE inversions}

\author{T. Felipe\inst{\ref{inst1},\ref{inst2}}
\and 
H. Socas-Navarro\inst{\ref{inst1},\ref{inst2}}
}


\institute{Instituto de Astrof\'{\i}sica de Canarias, 38205, C/ V\'{\i}a L{\'a}ctea, s/n, La Laguna, Tenerife, Spain\label{inst1}
\and 
Departamento de Astrof\'{\i}sica, Universidad de La Laguna, 38205, La Laguna, Tenerife, Spain\label{inst2} 
}

\abstract
{Spectropolarimetric inversions are a fundamental tool to diagnose the solar atmosphere. Chromospheric inferences rely on the interpretation of spectral lines that are formed under Non Local Thermodynamic Equilibrium (NLTE) conditions. In the presence of oscillations, changes in the opacity impact the response height of the spectral lines and hinder the determination of the real properties of the fluctuations.} 
{We aim to explore the relationship between the chromospheric oscillations inferred by NLTE inversion codes and the intrinsic fluctuations in velocity and temperature produced by the waves.}
{Numerical simulations of wave propagation in a sunspot umbra have been computed with the code MANCHA. The NLTE synthesis and inversion code NICOLE has been used to compute spectropolarimetric \CaII\ 8542 \AA\ line profiles for the atmospheric models obtained as the output from the simulations. The synthetic profiles have been inverted and the inferences from the inversions have been compared with the known atmospheres from the simulations.}
{NLTE inversions of the \CaII\ 8542 \AA\ line capture low frequency oscillations, including those in the main band of chromospheric oscillations around 6 mHz. In contrast, waves with frequencies above 9 mHz are poorly characterized by the inversion results. Velocity oscillations at those higher frequencies exhibit clear insights of opacity fluctuations since the power of the signal at constant optical depth greatly depart from the power of the oscillations at constant geometrical height. The main response of the line to velocity fluctuations comes from low chromospheric heights, whereas the response to temperature shows sudden jumps between the high photosphere and the low chromosphere. This strong variation in the heights where the line is sensitive to temperature is revealed as a strong oscillatory power in the inferred fluctuations, much stronger than the actual power from the intrinsic temperature oscillations.}
{Our results validate the use of NLTE inversions to study chromospheric oscillations with frequencies below $\sim$9 mHz. However, the interpretation of higher frequency oscillations and the power of temperature oscillations must be addressed with care since they exhibit signatures of opacity oscillations. }


\maketitle


\section{Introduction} \label{sect:intro}
The study of the solar atmosphere heavily relies on the observation and interpretation of the solar spectra, often not only using spectroscopic data but also full polarimetry. The \CaII\ 8542 \AA\ line is one of the most exploited spectral lines for probing the solar chromosphere. The interpretation of its spectral profiles requires NLTE diagnostics. Several NLTE inversion codes have been developed with this aim, such as NICOLE \citep{SocasNavarro+etal2015}, STiC \citep{delaCruz-Rodriguez+etal2019}, SNAPI \citep{Milic+vanNoort2018}, and DeSIRe \citep{Ruiz-Cobo+etal2022}. Due to the large amount of computational resources required by these inversions, pioneering studies using these tools were mostly restricted to analyzing a few spectral profiles \citep[\eg,][]{SocasNavarro+etal2000a} or spatially coherent maps with a reduced resolution for a few time steps \citep{delaCruz-Rodriguez+etal2013}. Until recently, chromospheric oscillations have been out of the scope of the works using NLTE inversions since they require the analysis of long temporal series with high temporal cadence and, thus, the inversion of numerous spectral profiles. Several works have employed alternative techniques to derive the chromospheric plasma properties from the interpretation of the \CaII\ 8542 \AA\ line, such as the lambdameter \citep[\eg,][]{Chae+etal2018} or bi-sector \citep[\eg,][]{Grant+etal2022} methods to infer the velocity. However, in umbral regions, these methods are challenged by the common display of emission near the core of the line as a manifestation of umbral flashes. The interpretation of these profiles requires sophisticated analysis tools, like inversion codes. 

Thanks to the improvement of the computational capabilities, recent works have been able to perform more comprehensive studies of sunspot chromospheres using NLTE inversions of the \CaII\ 8542 \AA\ line \citep{Henriques+etal2017,Joshi+delaCruzRodriguez2018, Henriques+etal2020, Houston+etal2020}. Also, new methods based on machine learning techniques are being developed to diagnose the solar chromosphere in a fast and computationally efficient way \citep{Vicente-Arevalo+etal2022}. The availability of physical information from larger maps and longer temporal series will enable the study of their oscillations in the common ground of Fourier and/or wavelet analyses. 

In the dynamic solar atmosphere, the contribution of different atmospheric heights to the formation of a spectral line changes with time \citep{Uitenbroek2003}. These variations are especially troubling for the study of solar oscillations. Spurious oscillations produced by the change in the formation height of the spectral lines in atmospheres with vertical gradients (known as opacity effects) can overlap the intrinsic fluctuations due to wave propagation. Disentangling the intrinsic oscillations from those produced by the opacity effect is thus fundamental for a proper interpretation of wave phenomena. 

Many reported magnetic field fluctuations in sunspots have been interpreted as a result of opacity effects \citep{BellotRubio+etal2000, Ruedi+Cally2003, Khomenko+etal2003}. Active regions are known to harbor vertical magnetic field gradients \citep[see][for a review]{Solanki2003, Borrero+Ichimoto2011}. In these atmospheres, the periodical displacements of the formation region of spectral lines introduce spurious oscillations in the inferred magnetic field. In contrast, the opacity effect is barely discussed in the examination of oscillations in other quantities, such as temperature and velocity. Indeed, the lower solar atmosphere exhibits significant vertical gradients in temperature. Velocity gradients are also expected since the amplitude of the oscillations increases with height due to the drop of the density \citep{Centeno+etal2006}. Both vertical gradients will certainly leave an imprint in the temperature/velocity fluctuations measured from any spectral line formed at those heights.

In this study, we focus on the umbral oscillations inferred from the analysis of the \CaII\ 8542 \AA\ line. This spectral line is sensitive to a broad range of heights, from the photosphere at the wings to the chromosphere at the core of the line \citep{Cauzzi+etal2008}. Nowadays, it is one of the most employed lines for the study of the solar chromosphere \citep[\eg,][]{SocasNavarro2005, Kleint2012, RouppevanderVoort+delaCruzRodriguez2013, Kuridze+etal2018, Murabito+etal2019}. It is also commonly used for the inspection of umbral oscillations and, more specifically, the development of umbral flashes \citep{SocasNavarro+etal2000a, delaCruz-Rodriguez+etal2013,Henriques+etal2017,Houston+etal2018,Bose+etal2019,Houston+etal2020}. 

Variations in the geometrical heights where the \CaII\ 8542 \AA\ line is sensitive during umbral flashes have been reported from the analysis of observations \citep{Joshi+delaCruzRodriguez2018} and numerical modeling \citep{Felipe+etal2021a}. Here, we address how how this opacity effect impacts the chromospheric umbral oscillations inferred with the \CaII\ 8542 \AA\ line. We have constructed synthetic Stokes profiles from the atmospheres computed with numerical simulations and then inverted those profiles. The numerical methods are briefly described in Sect. \ref{sect:methods} and the response functions of the line are introduced in Sect. \ref{sect:RFs}. In Sects. \ref{sect:velocity} and \ref{sect:temperature} we present the results obtained for velocity and temperature oscillations, respectively. Finally, conclusions are discussed in Sect. \ref{sect:conclusions}.


\begin{figure}[!ht] 
 \centering
 \includegraphics[width=9cm]{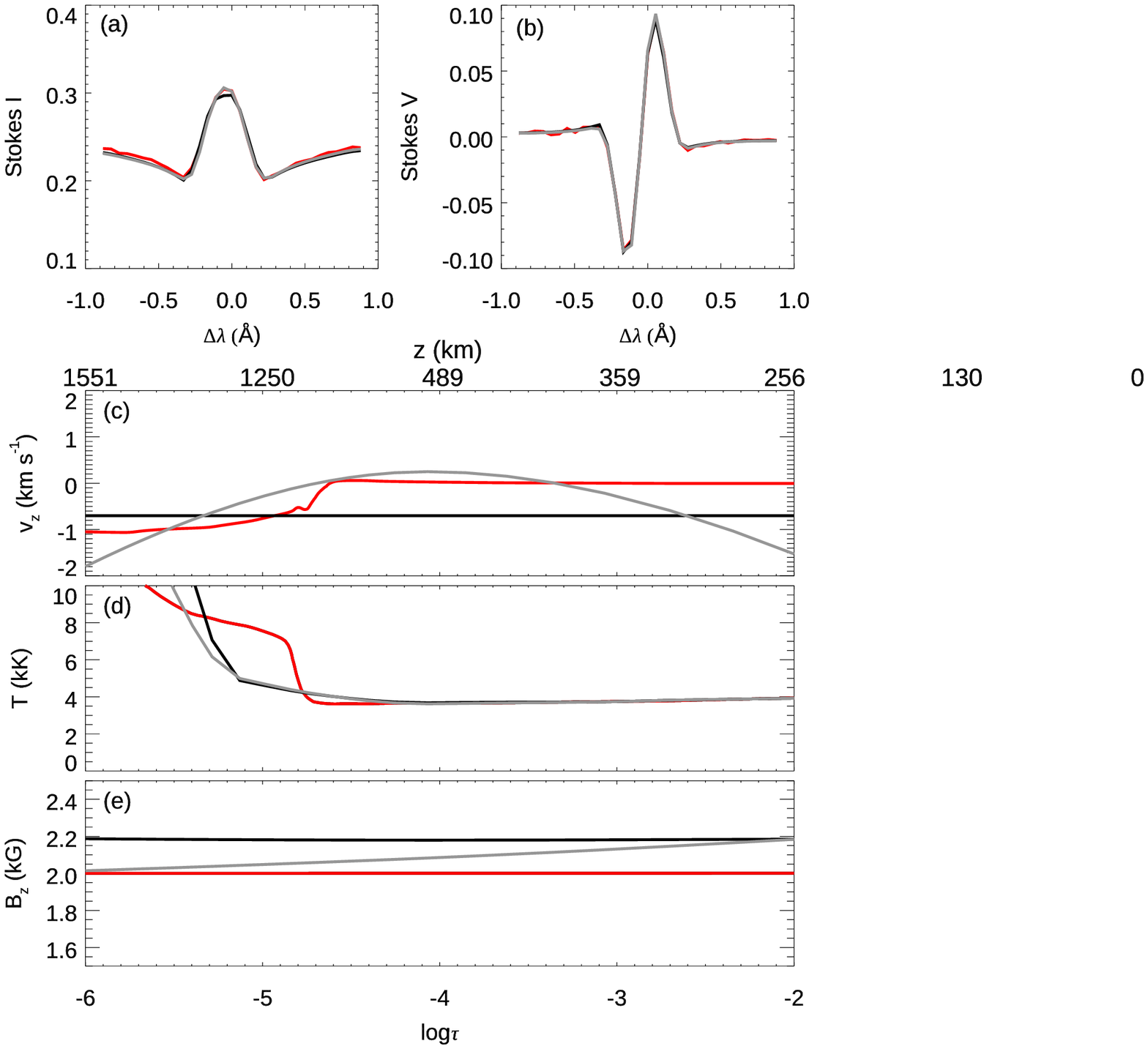}
  \caption{Stokes profiles and atmospheric models from an umbral flash. Top panels: Stokes I (panel a) and Stokes V (panel b) profiles. Bottom panels: vertical stratification of the velocity (panel c), temperature (d), and magnetic field (e) as a function of the optical depth (bottom axis) and geometrical height (top axis). In all panels, the red lines illustrated the quantities directly obtained from the numerical simulation, whereas black and grey lines are the results obtained from inversions with one or three velocity nodes, respectively.}      
  \label{fig:inversions}
\end{figure}

\section{Numerical methods} \label{sect:methods}

We have analyzed numerical simulations of wave propagation in a sunspot umbra computed with the code MANCHA \citep{Khomenko+Collados2006, Felipe+etal2010a}. These are the same simulations previously studied in \citet{Felipe+etal2021a, Felipe+etal2021b}. In those works, we describe the numerical setup, the calculation of the synthetic \CaII\ 8542 \AA\ spectropolarimetric profiles, and the inversion of those synthetic profiles \citep[only in][]{Felipe+etal2021b}. We refer the reader to those publications for a detailed description of the numerical methods. For completeness, in this manuscript we briefly describe them.

The numerical code computes the evolution of the perturbations to a background model. This model is a modified \citet{Avrett1981} umbral model, which was extended to the solar interior and corona. Simulations were computed using the 2.5D approximation (two-dimensional domain but keeping the three coordinates from vectors). The vertical domain spans from $z=-1.14$ Mm to $z=3.50$ Mm, with a constant vertical step of 10 km. In the horizontal dimension, we set the same background stratification at all spatial positions (96 points with a horizontal spatial step of 50 km), and periodic boundary conditions were established. The background model is permeated by a constant vertical magnetic field with a strength of 2000 G.

Waves are excited by a driver that reproduces actual umbral oscillations, including typical photospheric and chromospheric velocity amplitudes and power spectra. The spatial and temporal evolution of the driver were retrieved from photospheric umbral observations acquired with a slit spectrograph in the \SiI\ 10827 \AA\ line \citep{Felipe+etal2018b}. The driver was introduced as a vertical force directly implemented in the equations, following \citet{Felipe+etal2011} and \citet{Felipe+Sangeetha2020}. It changes along the horizontal dimension of the computational domain, corresponding to the direction along the slit of the spectrograph. This setup introduces horizontal variations in a simulation that otherwise would be purely one-dimensional (the magnetic field is constant and the stratification of the background is the same for all spatial positions). Instead, a two-dimensional simulation was employed to improve the statistical significance of the results since it allows us to sample a larger number of spectral profiles (including the synthesis and inversion of numerous umbral flashes with various properties) and to compute the spatial average of some of the quantities of interest, such as power spectra.

Synthetic spectropolarimetric \CaII\ 8542 \AA\ profiles were computed by feeding the NLTE code NICOLE \citep{SocasNavarro+etal2015} with the output from the simulation. The spectral resolution has been degraded by convolving the high-resolution profiles with Gaussians with an FWHM of 100 m\AA, obtaining a spectral step of 55 m\AA\ \citep[approximately the Nyquist frequency of the CRISP instrument at 8542 \AA;][]{Scharmer+etal2008}. Random noise has been added to produce profiles with a signal-to-noise of $1\times 10^{-3}$ in units of continuum intensity, comparable to that obtained in actual umbral flash observations \citep[\eg,][]{delaCruz-Rodriguez+etal2013}.

The central part of the umbra (46 points spanning 2.3 Mm in the horizontal direction) have been inverted for the whole temporal series (55 min of simulations/syntheses with a cadence of 5 s). A total of 30,360 profiles have been inverted. The NICOLE code was also employed to carry out these inversions. Two independent inversions of the whole set were performed. All of them employed a single cycle, with 6 nodes in temperature and 3 nodes in vertical magnetic field. Since the simulation has a purely vertical magnetic field and Stokes $Q$ and $U$ exhibit very weak signals, we did not invert the transversal magnetic field. Three velocity nodes were selected in one of the inversion sets, while in the other a single velocity node was imposed. 

An example of the simulated atmospheric models, their corresponding synthetic profiles, and the outcome from the inversions during an umbral flash is illustrated in Figure \ref{fig:inversions}. Both inversion setups (with a different number of velocity nodes) provide a good fit of the Stokes profiles and a fair characterization of the atmospheric stratification. The inversion with one velocity node (black lines) captures the actual velocity around $\log\tau=-5$, where the sensitivity of the \CaII\ 8542 \AA\ line to velocity is maximum during umbral flashes (see Section \ref{sect:velocity_formation}). The inferred chromospheric magnetic field exhibits a discrepancy of almost 200 G between both inversions. See \citet{Felipe+etal2021b} for a discussion of the limitations of the \CaII\ 8542 \AA\ line to measure magnetic fields.

\begin{figure}[!ht] 
 \centering
 \includegraphics[width=9cm]{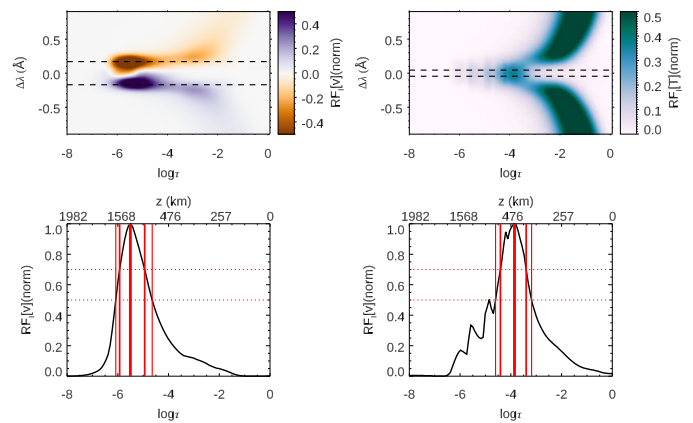}
  \caption{Response functions of the \CaII\ 8542 \AA\ intensity to line-of-sight velocity (left panels) and temperature (right panels) at a randomly chosen time step and spatial position. Top panels show the dependence of the normalized response function with wavelength and optical depth. Bottom panels illustrates the response functions integrated in the core of the line. The spectral region employed for the integral is delimited by the horizontal dashed lines in the top panels. Vertical red lines with different thickness indicate the optical depth (bottom axis) and geometrical height (top axis) where the normalized integrated response function is maximum (thicker line), above 0.7 (middle thick lines), and above 0.5 (thinner lines).}      
  \label{fig:RFs}
\end{figure}

\begin{figure}[!ht] 
 \centering
 \includegraphics[width=9cm]{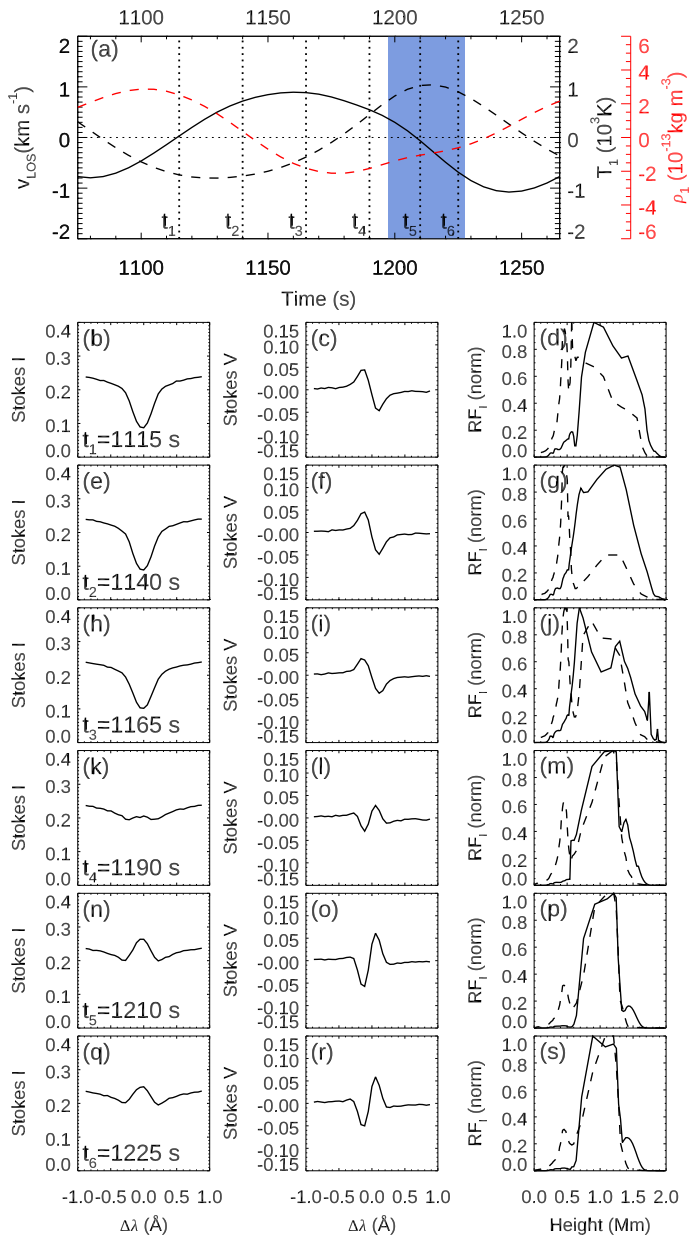}
  \caption{Temporal evolution of a synthetic umbral flash in \CaII\ 8542 \AA. Panel a: Vertical velocity (black solid line, left axis), temperature perturbation (black dashed line, right axis), and density perturbation (red dashed line, red right axis) as a function of time. The blue-shaded area denotes the times when the core of the line is in emission. Vertical dotted lines indicate the time steps plotted in panels b-s. Bottom panels: Stokes I (left column), Stokes V (middle column), and response functions of the intensity to velocity (solid line) and temperature (dashed line) as a function of geometrical height (right column). Each row correspond to the time shown in the left column and indicated by vertical dotted lines in panel a.}      
  \label{fig:Stokes_RFs}
\end{figure}

\section{Temporal evolution of \CaII\ 8542 \AA\ response functions} \label{sect:RFs}

Umbral chromospheres experience remarkable changes during the passage of waves. Fluctuations in velocity, temperature, and density modify the atmospheric stratification, with strong implications for radiative transfer. The imprint of these fluctuations is clearly visible in the Stokes profiles, which develop into umbral flashes. They also produce variations in the optical depth, source function, and opacity, which must be taken into account to understand the output radiation. In this work, we characterize the contribution of the different atmospheric layers to the measured spectra by computing the response functions of the intensity. 

Figure \ref{fig:RFs} illustrates the response functions of \CaII\ 8542 \AA\ intensity to velocity and temperature for a case when the core of the line is in absorption. A positive (negative) value of the response function indicates that an increase in the velocity/temperature at that optical depth will produce an increase (reduction) of the intensity at the corresponding wavelength of the line profile. In the illustrated example, the main contribution of the velocity to the line profile is concentrated at chromospheric heights. The two lobes with opposite signs indicate the intensity changes associated with the Doppler shift. A positive (downward) velocity shifts the line core to higher wavelengths and, thus, the intensity will increase at one side of the line and will decrease and the other side. In this work, we focus on the study of the chromospheric oscillations that can be measured with the \CaII\ 8542 \AA\ line. In the following, we will restrict the discussion of the response functions to those computed for the core of the spectral line. With this aim, we have integrated the absolute value of the response functions in the central part, in the range $\pm 17.5$ m\AA\ (for velocity) and $\pm 7.5$ m\AA\ (for temperature) from the line center. Bottom panels from Figure \ref{fig:RFs} show the integrated response functions.

Figure \ref{fig:Stokes_RFs} illustrates an umbral flash event, including the Stokes $I$ and $V$ profiles at certain time steps and their corresponding intensity response functions to temperature and velocity. The response functions are plotted as a function of geometrical height. They are computed from the simulated models, where we have access to an accurate stratification in geometrical scale. This situation differs from the analysis of observations, where the inversions return the stratification in optical depth and the geometrical height is calculated assuming hydrostatic equilibrium. The validity of this assumption in the case of umbral flashes is limited since they take place in a magnetized atmosphere and are associated with strong plasma flows. Our approach is not affected by this limitation.

The top panel of Figure \ref{fig:Stokes_RFs} shows the chromospheric temporal evolution at a randomly chosen position during one period (around three minutes). During this time, an umbral flash is developed, as seen in the appearance of the intensity emission core (left column) and the reversal of Stokes V (middle column). The core-integrated response functions (right column) exhibit remarkable variations during the different phases of the umbral flash. Initially, when the atmosphere is approximately at rest ($t_{1}=1115$ s), the response function to velocity (panel d) shows a high contribution from a broad range of heights (around 1 Mm wide), with the maximum at $z\sim 0.9$ Mm. In contrast, during the umbral flash (\eg, $t_{5}=1210$ s), the response function to velocity is narrower and the peak is shifted to $z\sim 1.2$ Mm. The core-integrated response function of the intensity to temperature exhibits even more striking variations than the response functions to velocity. Whereas during the umbral flash (last three rows) they are similar, in the initial stages the response function to temperature shows a peak at $z\sim 0.5$ Mm, indicating that at those time steps the information from Stokes $I$ contains a remarkable contribution from the high-photosphere. 

The results illustrated in Figure \ref{fig:Stokes_RFs} clearly show that the radiation measured in the core of the \CaII\ 8542 \AA\ line not only includes the imprint from a wide range of heights but also that those heights significantly change during the evolution of umbral flashes.

\begin{figure}[!ht] 
 \centering
 \includegraphics[width=9cm]{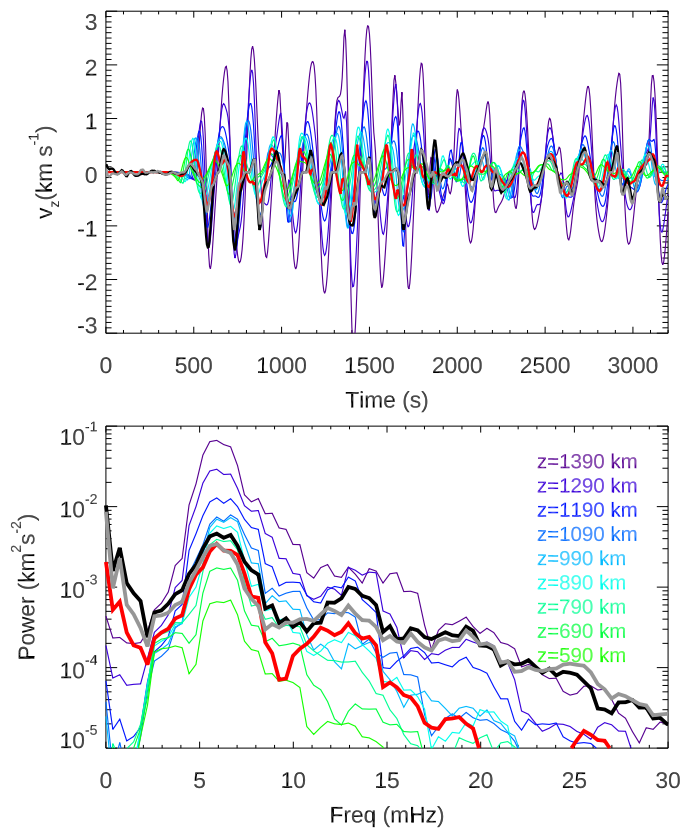}
  \caption{Comparison between velocity oscillations in geometrical scale, optical depth, and those inferred from the inversion of synthetic \CaII\ 8542 \AA\ profiles. Top panel: temporal evolution of the vertical velocity at a randomly chosen spatial position. Bottom panel: Power spectra of the vertical velocity averaged for all the spatial position from the numerical simulation. In both panels, the black lines correspond to the velocity inferred from inversions of the synthetic profiles with one node in velocity, the grey lines indicate the velocity inferred from inversions with 3 nodes in velocity averaged in $\log\tau=[-4.8,-5.4]$, and the red lines correspond to the actual velocity from the simulations averaged in the same range of optical depths. Color lines with a gradient green-blue-violet represent the velocity (top panel) and power spectra (bottom panel) at constant geometrical height. They span from $z=590$ km to $z=1390$ km, which is approximately the range of geometrical heights where the \CaII\ 8542 \AA\ line is sensitive to the velocity (Figure \ref{fig:Stokes_RFs}), with a step of 50 km between consecutive lines. The legend in the bottom panel indicates the heights corresponding to some colors, as a reference.}      
  \label{fig:v_power}
\end{figure}

\section{Velocity oscillations} \label{sect:velocity}

\subsection{Comparison of velocity signals at constant geometrical height and constant optical depth} \label{sect:velocity_power}

The velocity measured from the core of the \CaII\ 8542 \AA\ line samples atmospheric heights in the range $z\sim [0.6,1.5]$ Mm (Figure \ref{fig:Stokes_RFs}). Figure \ref{fig:v_power} illustrates a comparison between velocity oscillations at fixed geometrical heights in that range, from $z=0.59$ Mm (green) to $z=1.39$ Mm (violet), with velocity fluctuations at constant optical depth. For the latter, three different measurements are plotted: velocities obtained from the inversions of the synthetic profiles with one (black lines) or three (grey lines) velocity nodes, and the actual velocity directly obtained from the simulation (red lines). The illustrated signals at constant optical depth correspond to the average velocity between $\log\tau=-4.8$ and $\log\tau=-5.4$ since this range provides the main contribution to the core of the  \CaII\ 8542 \AA, as given by the examination of the response functions.   

The velocity inferred from the inversions captures fairly well the actual oscillations at constant optical depth (top panel from Figure \ref{fig:v_power}). Both inversions (which differ in the number of velocity nodes) show similar results, although the agreement with the actual simulated velocity of the inversion with three velocity nodes is slightly better (see the amplitude of the upflows (negative velocity) at $t\sim 600$ s and $t\sim 700$). However, there is a systematic shift in the maximum positive velocity (downflows) when the higher chromospheric layers (above $z\sim 1.1$ Mm) exhibit a high amplitude (wavefronts between $t=800$ s and $t=1900$ s). In those cases, the maximum positive velocity obtained from the inversions (black and grey lines) takes place in phase with the velocity at $z\sim 1.1$ Mm and lags the maximum velocity at constant optical depth from the simulation (red line). The velocity fluctuations at constant optical depth (both from inversions and actual simulated values) can hardly be associated with a single geometrical height. Instead, they exhibit contributions from different layers, which depend on the phase of the oscillation.

The power spectra (bottom panel from Figure \ref{fig:v_power}) show that \CaII\ 8542 \AA\ inversions provide an excellent characterization of the main frequency (period) of chromospheric oscillations. The power of that main peak is comparable to that from oscillations at $z\sim 0.75$ Mm. In contrast, discrepancies are found at higher frequencies. Oscillations at constant optical depth exhibit a secondary power peak at around 13 mHz. This peak is absent in the power peak of oscillations at constant geometrical height (only the power spectra at $z\sim 1.2$ Mm shows some hints of a power enhancement at that frequency). This power excess is not an artifact from the solution of the inversion problem, but it is already present in the oscillations at a constant optical depth directly extracted from the simulation (red line). This fact points to the change in the layers where the spectral line is sensitive as a key feature to interpret oscillations in the 10-15 mHz band. In the case of even higher frequencies, the inversions greatly depart from the actual simulated power. The small amplitude of these fluctuations is beyond the expected precision of the inversion results. 

\begin{figure}[!ht] 
 \centering
 \includegraphics[width=9cm]{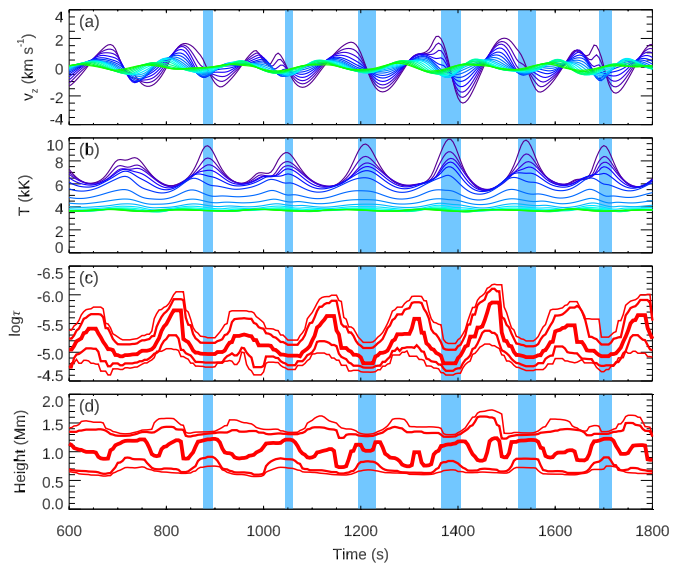}
  \caption{Oscillations in a randomly chosen location of the simulated umbra in the high chromosphere and low chromosphere (top two panels) and the regions where the \CaII\ 8542 \AA\ intensity is sensitive to the velocity (bottom two panels). Top panels show the temporal evolution of the vertical velocity (panel a) and temperature (middle panel) at constant geometrical heights. The colors have the same meaning as in Figure \ref{fig:v_power}. Bottom panels illustrate the range of optical depths (panel c) or geometrical heights (panel d) where the \CaII\ 8542 \AA\ intensity is sensitive to the velocity, as given by the examination of the response functions. The thickest lines indicate the layer where the response function is maximum. Thinner lines delimit successively the range of heights where the normalized response function is above 0.7, and 0.5. Blue-shaded regions indicate the times when the core of the line is in emission.}      
  \label{fig:evol}
\end{figure}

\subsection{Fluctuations in the formation height of the \CaII\ 8542 \AA\ line core} \label{sect:velocity_formation}
The response functions of intensity to velocity have been computed for all the atmospheres in the central part of the simulation, individually for each of the 660 time steps. Figure \ref{fig:evol} illustrates the variations in the optical depths (panel c) and geometrical heights (panel d) where the line core is sensitive to velocity. For each atmospheric model, we have integrated the response function in the central wavelengths and determined the height of the maximum response (thick red line) and the heights where the normalized response function exhibit some selected values (red lines with smaller thickness as lower response values are considered). The definition of these lines is illustrated in the bottom panels from Figure \ref{fig:RFs}.

The heights with higher sensitivity to the velocity exhibit periodic behavior, both in optical depth and geometrical height. During umbral flashes (blue-shaded areas, taking place at the times when the chromospheric temperature is maximum) the core of the \CaII\ 8542 \AA\ line is sensitive to a deeper optical depth (with the maximum response at around $\log\tau=-5.0$). Interestingly, this deeper optical depth is associated with high geometrical heights (around $z=1.2$ Mm). Generally, the geometrical height where the response of the core is maximum is maintained approximately constant during the entire umbral flash event. In contrast, during the quiescent phase of the oscillations, the core of the line is in absorption and the peak of its sensitivity is shifted to optical depths around $\log\tau=-5.7$ and geometrical heights as low as $z=0.7$ Mm.

\begin{figure}[!ht] 
 \centering
 \includegraphics[width=9cm]{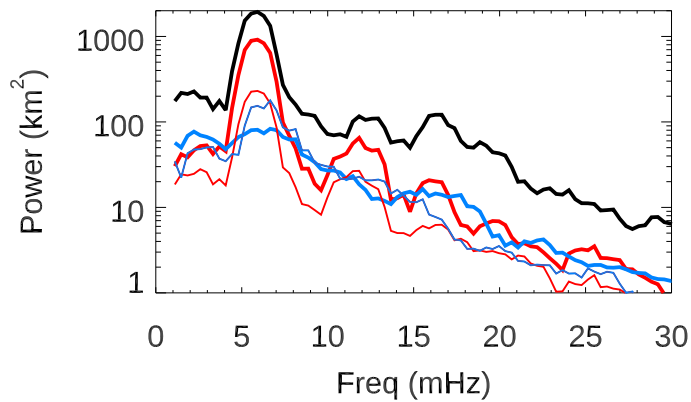}
  \caption{Power spectra of the formation height of the \CaII\ 8542 \AA\ line averaged for all locations inside the umbra. The lines represent the maximum of the response function (black line), the minimum height where the response function of the intensity to velocity is above a selected threshold (red lines), and the maximum height where the response function is above the threshold (blue lines). The thickness of the color lines indicates the threshold concerning the normalized response function, corresponding to 0.7, 0.5, and 0.3 from thicker to thinner. }      
  \label{fig:power_z}
\end{figure}

The periodicity of the fluctuations in the formation height of the line core has been evaluated by computing the average power spectra of the geometrical heights plotted in Figure \ref{fig:evol} that characterize the regions where the line is sensitive to velocity. Figure \ref{fig:power_z} shows that the main frequency of the oscillations is 6 mHz, in agreement with the oscillations in the 3-minute band widely reported from chromospheric umbral oscillations. Some secondary peaks appear at 10-16 mHz, especially for the peak of the response function (black line) and the lower height of the selected contours (red lines). These peaks coincide with the power enhancement found in the velocity inferred from the inversions (bottom panel from Figure \ref{fig:v_power}) and point to fluctuations in the sensitivity of the line core to height as the origin of the measured secondary power peak in velocity at 12-13 mHz (Figure \ref{fig:v_power}).

\begin{figure}[!ht] 
 \centering
 \includegraphics[width=9cm]{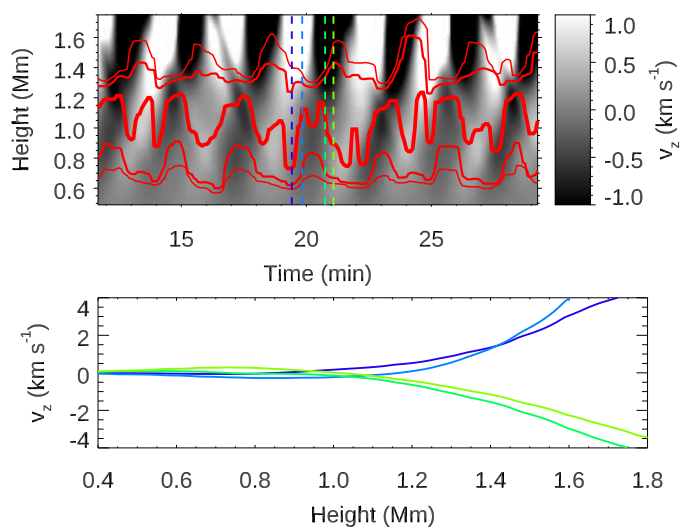}
  \caption{Top panel: Vertical velocity as a function of height (vertical axis) and time (horizontal axis) at a randomly chosen spatial position. Red lines have the same meaning as those in the bottom panel from Figure \ref{fig:evol}. Bottom panel: Vertical stratification of the velocity at some selected time steps. The color of the line indicates the corresponding time step, as given by the dashed vertical lines in the top panel. }      
  \label{fig:z_RFs}
\end{figure}

\subsection{Velocity gradients and opacity effects} \label{sect:velocity_gradients}
The effective formation height of the \CaII\ 8542 \AA\ line core significantly changes with the atmospheric variations produced by umbral oscillations. The presence of vertical gradients in the stratification of the atmosphere can thus lead to spurious oscillations in the quantities inferred from the interpretation of the spectral line. In addition to the intrinsic oscillations associated with wave travel, a second component produced by changes in the opacity can leave an imprint in the measurements.

Figure \ref{fig:z_RFs} illustrates the stratification of the vertical velocity between the high photosphere and the chromosphere during $\sim$18 min of simulation. In the top panel, the velocity (grey scale) is saturated outside the $\pm1$ km s$^{-1}$ range to better visualize its variations in the layers where the core of the line is formed (indicated by the red lines). The bottom panel shows the vertical velocity at some selected time steps. All of them correspond to times when the response function of the \CaII\ 8542 \AA\ line core is undergoing striking variations, with its maximum shifting from lower to higher layers (blueish lines) or vice versa (greenish lines). During those time steps, the atmospheric velocity exhibits strong vertical gradients. The combination of vertical gradients in the velocity with the variations in the layers where the line core is sensitive to velocity produces the aforementioned additional component in the measured velocity as a result of opacity changes.

\begin{figure}[!ht] 
 \centering
 \includegraphics[width=9cm]{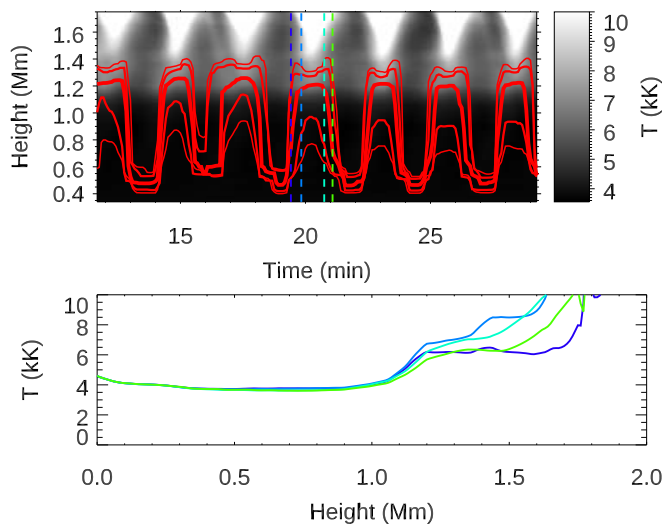}
  \caption{Top panel: Temperature as a function of height (vertical axis) and time (horizontal axis) at a randomly chosen spatial position. Red lines indicate the height of the peak of the response function of intensity to temperature (thickest line) and the range of heights where the normalized response function is above 0.7 (medium-thick line), and 0.5 (thinner line). Bottom panel: Vertical stratification of the temperature at some selected time steps. The color of the line indicates the corresponding time step, as given by the dashed vertical lines in the top panel. }      
  \label{fig:z_RFs_t}
\end{figure}

\begin{figure}[!ht] 
 \centering
 \includegraphics[width=9cm]{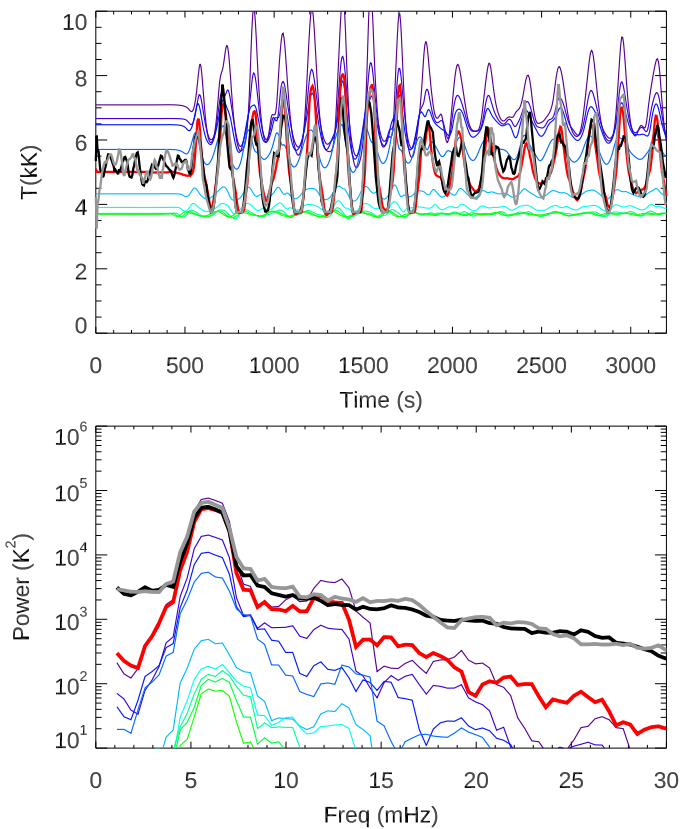}
  \caption{Comparison between temperature oscillations in geometrical scale, optical depth, and those inferred from the inversion of synthetic \CaII\ 8542 \AA\ profiles. Top panel: temporal evolution of the temperature at a randomly chosen spatial position. Bottom panel: Power spectra of the temperature averaged for all the spatial positions from the numerical simulation. The color lines have the same meaning as in Figure \ref{fig:v_power}. }      
  \label{fig:t_power}
\end{figure}

\section{Temperature oscillations} \label{sect:temperature}

Figure \ref{fig:Stokes_RFs} shows that the response function of the intensity to temperature also experiences striking variations during umbral oscillations. The range of heights where the temperature is probed by the core of the \CaII\ 8542 \AA\ line is even broader than that for the velocity, with the peak of the response function changing from below $z\sim 0.5$ Mm ($t_2=1140$ s) up to $z\sim 1.2$ Mm ($t_4=1190$ s). The top panel from Figure \ref{fig:z_RFs_t} shows the periodicity of these fluctuations. The region where the line core is sensitive to temperature exhibits sudden variations from the photosphere (when the temperature perturbation is low) to the chromosphere (at those times when the chromospheric temperature increases). In the umbral atmospheres computed during the simulation, the chromosphere is generally $\sim$2000 K hotter than the photosphere (bottom panel from \ref{fig:z_RFs_t}). This way, the intrinsic temperature enhancements produced by the waves will be accompanied by an additional temperature increase caused by the displacement in the region where the line is sensitive to temperature.

The comparison between the temperature oscillations at constant geometrical height and constant optical depth illustrated in Figure \ref{fig:t_power} shows clear indications of height variations in the origin of the temperature signals. During the first 500 s of simulation, when the fluctuations driven under the photosphere have not yet reached upper atmospheric layers and the chromosphere is approximately at rest, the temperature at constant optical depth (both from inversions and directly extracted from the simulation) is consistent with the temperature at $z\sim 0.89$ Mm (light-blue line). However, later times exhibit oscillations with amplitude significantly higher than that found for that geometrical height. When the chromospheric temperature is maximum (during umbral flashes), the temperature at constant optical depth is comparable to the temperature at $z\sim 1.2$ Mm (dark-blue/violet line). In contrast, the lowest values of the temperature at constant optical depth are similar to the temperature measured at $z\sim 0.5$ Mm (light green line). The temperature oscillations inferred from the inversions of the \CaII\ 8542 \AA\ line show a fair agreement with the real temperature oscillations directly extracted from the output of the simulation, although some discrepancies are found at the times when the temperature is maximum. 

The oscillations at constant optical depth have peak-to-peak amplitude around 2000 K, although the strongest wavefronts exhibit a change between the minimum and maximum temperature up to 4000 K. This amplitude is comparable to that from oscillations at $z\sim1.3$ Mm (violet line in Figure \ref{fig:t_power}). This is also illustrated by the power spectra (bottom panel from Figure \ref{fig:t_power}). The frequency of the main power peak at 6 mHz, present in oscillations at constant geometrical height and optical depth, is perfectly reproduced by the inversion results. Its power is similar to the power from oscillations at $z\sim1.3$ Mm, which is the upper limit of the region sampled by the \CaII\ 8542 \AA\ line core. In contrast, the power of the oscillations at $\sim0.89$ Mm (whose temperature is probed by the core of the line when the atmosphere is approximately at rest, first 500 s of simulation) is almost two orders of magnitude lower. This result indicates that opacity oscillations are a significant component of the total temperature variations measured during umbral flashes. Both intrinsic and opacity oscillations take place in phase, with the core of the line shifting to upper (hotter) layers when the temperature of the intrinsic oscillations increases. 

Similarly to the case of the velocity (Figure \ref{fig:v_power}), the power of temperature oscillations at constant optical depth also exhibits a secondary peak at around 12-13 mHz (red line in the bottom panel from Figure \ref{fig:t_power}). However, the strength of this power peak is lower than the fluctuations that can be captured by the inversions. 

\section{Discussion and conclusions} \label{sect:conclusions}

In this paper, we have characterized the relationship between the quantities inferred from inversions of the \CaII\ 8542 \AA\ line and the intrinsic oscillations at constant geometrical height. The former is affected by the opacity effect, which changes the response height of the spectral line and severely affects the interpretation of the results. This goal has been addressed through numerical simulations of wave propagation in an umbral model. The stratification of the simulated velocity and temperature fluctuations has been compared with the fluctuations of the same quantities at constant optical depth. In addition, synthetic \CaII\ 8542 \AA\ profiles have been computed from the output of the simulations, and the inferences from their inversions have also been critically evaluated. 

Our analysis focused on the examination of the chromospheric velocity and temperature fluctuations, obtained by averaging the atmosphere (both the actual simulated model and those inferred from the inversion of the synthetic profiles) in a range of optical depths where the \CaII\ 8542 \AA\ line is generally sensitive to changes in the atmosphere. Our results show that inversions of the \CaII\ 8542 \AA\ line provide a good characterization of the velocity oscillations in the low chromosphere. They capture the amplitude (and power) of the oscillations and the frequency of the main power peak at around 6 mHz, indicating that they provide a reliable estimation of the main oscillatory period. However, some discrepancies are also obvious. The time steps of the velocity maximum (downflow) inferred from the inversions lag those found in the simulations at the low chromosphere (around 60 s delay). Instead, they go in phase with the maximum velocity and higher mid-chromospheric layers (top panel from Fig. \ref{fig:v_power}). This points to a change in the optical depths where the \CaII\ 8542 \AA\ line is sensitive to the velocity during different phases of the oscillation. An examination of the evolution of the response function shows that during the downflowing phase in the mid-chromosphere ($z \in [1000,1400]$), the response of the line is shifted upward to the range $\log\tau \in [-5.0,-6.1]$, with the maximum of the response reaching up to $\log\tau=-5.9$ (Fig. \ref{fig:RFs}).

The velocity power spectrum inferred from the inversions with three velocity nodes closely matches the simulated spectrum (averaged in $\log\tau$) for frequencies lower than 8 mHz. For higher frequencies, the inferred power is significantly stronger than the actual simulated power. Interestingly, the simulated power averaged in $\log\tau$ exhibits a remarkable dip at around 9.5 mHz (bottom panel from Fig. \ref{fig:v_power}). This dip is absent in the power computed at constant geometrical heights. The dip and subsequent peak are partially captured by the inversions, which show a power excess between 12 and 14 mHz. The examination of the response functions shows that a similar power peak is found in the height where the line is sensitive to the velocity (Fig. \ref{fig:power_z}) and that changes in the response height take place when strong velocity gradients are present (Fig. \ref{fig:z_RFs_t}). The presence of power peaks (or dips) at the sunspot chromosphere has been suggested to be caused by the existence of a chromospheric resonant cavity \citep{Jess+etal2020, Felipe+etal2020} or by harmonics resulting from the non-linearity of the three-minute oscillations \citep{Chae+etal2018, Felipe2021, Chai+etal2022}. Our results suggest caution with the interpretation of power peaks with frequencies higher than 9 mHz in \CaII\ 8542 \AA\ observations.    

The opacity effect also has a remarkable impact on the measured temperature fluctuations. The region where the core of the \CaII\ 8542 \AA\ line exhibits a significant response to temperature oscillates between $z\sim 500$ km and $z\sim 1200$ km, that is, a region with strong vertical gradients in temperature (Fig. \ref{fig:z_RFs_t}). Temperature fluctuations produced by the opacity effect are in phase with the intrinsic temperature oscillations associated with wave propagation. When the temperature (at constant geometrical height) increases, the response of the line is shifted to upper (hotter) layers. This leads to stronger inferred temperature fluctuations, which exhibit a high power in the three-minute band (higher than the power in most of the geometrical heights where the core of the line forms). The frequency of this power peak is well captured by the inversions.

Our goals are in line with the recent work by \citet{Keys+etal2021}. They evaluated the inversion results of oscillations in the photospheric 6301 \AA\ and 6302 \AA\ lines using a similar approach of comparing the output from the inversions with known simulated atmospheres. Their analysis focused on short-period (24 s) waves, finding a good match between simulations and inversions except for some discrepancies during the passage of waves. These deviations are due to the small height range perturbed by the waves in comparison with the range where the lines are sensitive. Our results for chromospheric fluctuations inferred from a line formed under NLTE conditions indicate that oscillations with frequencies higher than 9 mHz (periods shorter than $\sim 110$ s) are hardly well characterized by the output of the inversions. Low-frequency acoustic waves (or, more precisely, slow magnetoacoustic waves) have a longer wavelength, which is comparable to the range of heights where the \CaII\ 8542 \AA\ line is formed. However, at chromospheric heights the temperature and velocity change at spatial scales much shorter than the photon free path, which makes it impossible to precisely determine their vertical stratification. The inversion can only be interpreted as the average properties in a region large enough to affect the emerging radiation. This way, our analysis has focused on the average chromospheric fluctuations. A comparison between the models inferred from the inversions and the actual vertical stratification is illustrated in Fig. \ref{fig:inversions}. This limitation could be lessened by performing multi-line inversions with several spectral lines that probe similar atmospheric layers.

\begin{acknowledgements}
Financial support from grants PGC2018-097611-A-I00 and PID2021-127487NB-I00, funded by MCIN/AEI/ 10.13039/501100011033 and by “ERDF A way of making Europe” is gratefully acknowledged. TF acknowledges grant RYC2020-030307-I funded by MCIN/AEI/ 10.13039/501100011033 and by “ESF Investing in your future”. We acknowledge the contribution of Teide High-Performance Computing facilities to the results of this research. TeideHPC facilities are provided by the Instituto Tecnol\'ogico y de Energ\'ias Renovables (ITER, SA). URL: \url{http://teidehpc.iter.es}. 
\end{acknowledgements}

\bibliographystyle{aa} 
\bibliography{biblio.bib}

\end{document}